\documentclass[twocolumn,showpacs,preprintnumbers,amsmath,amssymb]{revtex4}

\usepackage{graphicx}
\usepackage{dcolumn}
\usepackage{bm}
\newcommand{\ps}{p\hspace{-0.44em}/\hspace{0.06em}}
\newcommand{\eq}[1]{Eq.~(\ref{#1})}
\newcommand{\fig}[1]{Fig.~\ref{#1}}

\newcommand{\bbmd}{\bbd\ mixing}

\newcommand{\bbd}{\ensuremath{B_d\!-\!\Bbar{}_d\,}}

\newcommand{\Bbar}{\,\overline{\!B}}

\newcommand{\real}{\mathrm{Re}\,}

\begin{document}

\preprint{TTP09-24}

\title{Effects of right-handed charged currents\\ 
       on the determinations of $|V_{ub}|$ and $|V_{cb}|$}
%
\author{Andreas Crivellin}
\affiliation{Institut f\"ur Theoretische Teilchenphysik\\
               Karlsruhe Institute of Technology, 
               Universit\"at Karlsruhe, \\ D-76128 Karlsruhe, Germany}%
\date{July 14, 2009; revised November 10, 2009}
\begin{abstract}
  We study the effect of a right-handed coupling of quarks to the
  W-boson on the measurements of $|V_{ub}|$ and $|V_{cb }|$. It is shown that
  such a coupling can remove the discrepancies between the
  determinations of $|V_{ub}|$ from $B\to \pi \ell \nu$, $B^+\to \tau^+
  \nu$ and $B\to X_u \ell \nu$.  Further the measurements of $|V_{cb}|$
  from $B\to D^* \ell \nu$, $B\to D \ell \nu$ and inclusive $B\to X_c
  \ell\nu$ decays can be brought into better agreement.  We demonstrate
  that a right-handed coupling can be generated within the MSSM by a
  finite gluino-squark loop. The effect involves the parameters
  $\delta^{u\;RL}_{23}$ and $\delta^{u\;RL}_{13}$ of the squark mass
  matrices, which are poorly constrained from other processes. On the
  other hand, all gluino-squark corrections to the regular left-handed
  coupling of the W-boson are found to be too small to be relevant.
  \end{abstract}
\maketitle

\section{\label{sec:level1}Introduction}

In the standard model (SM) with its gauge group $\rm{SU(3)_C \times SU(2)_L\times U(1)}$ the tree-level W coupling has a pure $V-A$
structure meaning that all charged currents are left-handed.
Right-handed charged currents were first studied in the context of
left-right symmetric models \cite{Senjanovic:1975rk} which enlarge the
gauge group by an additional $SU(2)_R$ symmetry between right-handed
doublets. In these models new right-handed gauge bosons $W_R$, $Z_R$
appear and the physical SM-like W-boson has a dominant left-handed
component with a small admixture of $W_R$. The latter will generically
lead to small right-handed couplings to both quarks and leptons.  The
right-handed mass scale inferred from today's knowledge on neutrino
masses is so large that all right-handed gauge couplings are
undetectable. Most of these couplings are further experimentally
strongly constrained \cite{Amsler:2008zzb}.  A different source of
right-handed couplings of quarks to the W-boson can be loop effects,
which generate a dimension-6 quark-quark-W vertex.  In this case no
right-handed lepton couplings occur, as long as the neutrinos are
assumed left-handed. A generic analysis of such higher-dimensional
right-handed couplings has been studied in Ref.~\cite{Bernard:2007cf}
aiming at a better understanding of $K\to \pi \mu \nu$ data. The general effect of left- and right-handed anomalous couplings of the W to charm was studies in Ref.~\cite{He:2009hz}. The authors conclude that only the real part of the right-handed charm-bottom coupling can be sizable. The coupling of the W to up has be studied in \cite{Chen:2008se}.

We will investigate the effect of a right-handed W-coupling on the
extraction of $|V_{ub}|$ and $|V_{cb}|$ in section II and show that current
tensions between SM and data can be removed. In section III we will
calculate the loop-corrected W-coupling in the generic Minimal
Supersymmetric Standard Model (MSSM). We find that the right-handed
W-coupling can be as large as 20\%\ and brings the different
determinations of $|V_{ub}|$ into perfect agreement. The effect on
$|V_{cb}|$ is at most around 2\%, which alleviates the tension studied
in Sec.~II. 
Finally we conclude.

\section{Right-handed W couplings}
An appropriate framework for our analysis is an effective Lagrangian. 
Following the notation of Ref.~\cite{Grzadkowski:2008mf}, we write
\begin{equation}
{\cal L} = {\cal L}_{\rm SM} 
+ \frac{1}{\Lambda  } \sum_i C_i^{(5)} Q_i^{(5)} 
+ \frac{1}{\Lambda^2} \sum_i C_i^{(6)} Q_i^{(6)} 
+ {\cal O}\left(\frac{1}{\Lambda^3}\right),
\label{Leff}
\end{equation}
here ${\cal L}_{\rm SM}$ is the standard model (SM) Lagrangian, while
$Q_i^{(n)}$ stand for dimension-$n$ operators built out of the SM fields
and are invariant under the SM gauge symmetries. Such an effective theory approach is
appropriate for any SM extension in which all new particles are
sufficiently heavy ($M_{\rm new} \sim \Lambda \gg m_t$). As long as only
processes with momentum scales $\mu \ll \Lambda$ are considered, all heavy
degrees of freedom can be eliminated \cite{Appelquist:1974tg}, leading to the
effective theory defined in (\ref{Leff}). The operators $Q_i^{(5)}$ and $Q_i^{(6)}$ have been completely classified in Ref.~\cite{Buchmuller:1985jz}. Since $Q_i^{(5)}$ involve no quark
fields, they are not needed for our further discussion, and we skip the superscripts
``(6)'' at the dimension-six operators and the associated Wilson
coefficients $C_i$. In this article, we need the following dimension-six
operator describing anomalous right-handed W-couplings to quarks:
\begin{equation}
  Q_{RR}=\bar{u}_f \gamma^\mu P_R d_i 
  \left(\tilde \phi ^\dagger i D^\mu \phi\right)+h.c.
\label{Operator}
\end{equation}
where $\phi$ denotes the Higgs doublet and $\widetilde{\phi} =
i\tau^2\phi^*$. The Feynman rule for the $W$-$u_f$-$d_i$ interaction vertex,
\begin{equation}
\frac{ - i g_2 \gamma^\mu}{\sqrt 2 }
 \left( {V_{fi}^L P_L  + V_{fi}^R P_R } \right)
, \label{W-coupling}
\end{equation}
is found by combining the usual SM interaction with the extra
contributions that are obtained by setting the Higgs field in
\eq{Operator} to its vacuum expectation value.  In \eq{W-coupling}
$V^L_{fi}$ and $V^R_{fi}$ denote elements of the effective CKM matrices, which are not
necessarily unitary.  $V_{fi}^R$ is related to the Wilson coefficient in
\eq{Leff} via $V_{fi}^R = \frac{C_{RR}}{2 \sqrt{2} G_F \Lambda^2}$.
$V_{fi}^L$ receives contributions from the tree-level CKM matrix and 
the LL analogue of $Q_{RR}$ in \eq{Operator}. 

Right-handed couplings to light quarks have been studied in
Ref.~\cite{Bernard:2007cf} and to charm (up) quarks in Ref.~\cite{He:2009hz} (Ref.~\cite{Chen:2008se}). Ref. \cite{Dassinger:2008as} examines such
couplings in inclusive b$\to$c transitions. In
Ref.~\cite{Grzadkowski:2008mf} it was pointed out that very strong
constraints can be obtained on $V_{tb}^R$ from $b\to s\gamma$, because
the usual helicity suppression factor of $m_b/M_W$ is absent in the
right-handed contribution.  By the same argument $V_{ts}^R$ (or
$V_{td}^R$ if one considers $b\to d\gamma$) is tightly constrained.
Large effects concerning transitions between the first two generations
are unlikely, because $V_{us}^L$ and $V_{cd}^L$ are larger than other
off-diagonal CKM elements. Further deviations from Minimal Flavour
Violation (as defined in \cite{D'Ambrosio:2002ex}), i.e.\ deviations from Yukawa-driven flavour
transitions, are unlikely in the first two generations, but plausible
with respect to transitions involving the third generation
\cite{Crivellin:2008mq}.  We therefore focus our attention 
on the remaining two elements $V_{ub}^R$ and $V_{cb}^R$.

\subsection{Determination of $V_{ub}^L$ and $V_{cb}^L$}

\begin{figure}
\includegraphics[width=0.47\textwidth]{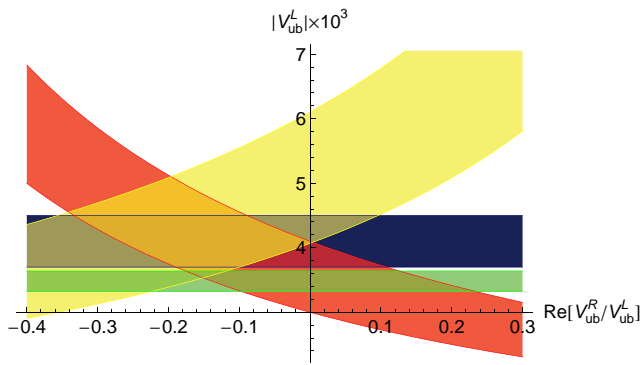}
\caption{\label{Vub} $\left|V^L_{ub}\right|$ as a function of
  $\rm{Re}\left[V^R_{ub}/V^L_{ub}\right]$ extracted from different
  processes. Blue(darkest): inclusive decays. Red(gray): $B\to \pi l\nu$.
  Yellow(lightest gray): $B\to \tau\nu$. Green(light gray): $V^L_{ub}$ determined from CKM unitarity.}
\end{figure}The experimental determination of $|V_{ub}|$ and $|V_{cb}|$ from both
inclusive and exclusive $B$ decays is a mature field by now
\cite{Amsler:2008zzb}. E.g. the form factors needed for $B\to \pi l\nu$
are known to 12\% accuracy \cite{Arnesen:2005ez}. More recently, also the leptonic decay $B\to
\tau \nu_\tau$ is studied in the context of $V_{ub}$.  To discuss the
impact of right-handed currents we denote the CKM element extracted from
data with SM formula by $V_{qb}$, where $q=u$ or $q=c$. If the matrix
element of a considered exclusive process is proportional to the vector
current, $V_{qb}^L$ and $V_{qb}^R$ enter with the same sign and the
"true" value of $V_{qb}^L$ in the presence of $V_{qb}^R$ is given by:
\begin{equation}
V_{qb}^L=V_{qb}-V_{qb}^R \label{truel}
\end{equation}
For processes proportional to the axial-vector current $V_{qb}^R$ enters
with the opposite sign as $V_{qb}^L$, so that
\begin{equation}
V_{qb}^L=V_{qb}+V_{qb}^R . \label{truer}
\end{equation}

In inclusive decays the interference term between the left-handed and
right-handed contributions is suppressed by a factor of $m_q/m_b$, so
that it is irrelevant in the case of $V_{ub}$ and somewhat suppressed in
the case of $V_{cb}$. The remaining dependence on $V_{qb}^R$ is
quadratic and therefore negligible. 

Starting with $|V_{ub}|$, we note that the determinations from inclusive
and exclusive semileptonic decays agree within their errors, but the
agreement is not perfect \cite{Amsler:2008zzb,Charles:2004jd}. The
analysis of $B\to\tau\nu$ is affected by the uncertainty in the
decay constant $f_B$. Within errors the three determination of
$|V_{ub}|$ are compatible, as one can read off from \fig{Vub}.  The
picture looks very different once the information from a global fit to
the unitarity triangle (UT) is included: As pointed out first by the
CKMFitter group, the measured value of $B\to\tau\nu$ suffers from a
tension with the SM of 2.4--2.7$\sigma$ \cite{Charles:2004jd}.  First,
the global UT fit gives a much smaller error on $|V_{ub}|$ (as a
consequence of the well-measured UT angle $\beta$); the corresponding
value is also shown in \fig{Vub}.  Second, the data on \bbmd\ exclude
very large values for $f_B$, which in turn cut out the lower part of the
yellow (light gray) region in \fig{Vub}. Essentially we realize from
\fig{Vub} that we can remove this tension while simultaneously bringing
the determinations of $|V_{ub}|$ from inclusive and exclusive
semileptonic decays into even better agreement. For this the
right-handed component must be around $\real (V_{ub}^R/V_{ub}^L) \approx
-0.15$. Since new physics may as well affect the other quantities
entering the UT, a more quantitative statement requires the
consideration of a definite model.

\begin{figure}[t]
\includegraphics[width=0.47\textwidth]{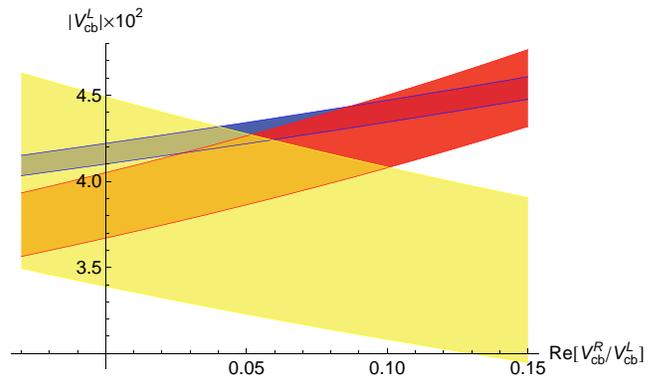}
\caption{\label{Vcb} $\left|V^L_{cb}\right|$ as a function of
  $\rm{Re}\left[V^R_{cb}/V^L_{cb}\right]$ extracted from different
  processes. Blue(darkest): inclusive decays. Red(gray): $B\to D^*l\nu$.
  Yellow(light gray): $B\to D l\nu$.}
\end{figure}

Next we turn to $|V_{cb}|$: The relative uncertainties in
the exclusive decays $B\to D^{*} l\nu$ and $B\to D l\nu$ and in the inclusive $B\to
X_c \ell \nu$ analyses are much smaller than in the $b\to u$ decays
considered above. Note that $B\to D l\nu$ only involves the
vector current so that \eq{truel} applies. $B\to D^{*} l\nu$ receives
contributions from both vector and axial vector currents, but
the contribution from the vector current is suppressed in
the kinematic endpoint region used for the extraction of $|V_{cb}|$.
Therefore \eq{truer} applies to $B\to D^{*} l\nu$. The impact of a
right-handed current on $B\to X_c \ell \nu$ has
been calculated in Ref.~\cite{Dassinger:2008as}.  \fig{Vcb} shows that the
agreement among the three values of $|V_{cb}|$ obtained from these decay
modes is not totally satisfactory within the SM.  One further realizes
that we can reduce the discrepancy to less than $1\sigma$ if a right-handed
coupling in the range $0.03\leq \real[V_{cb}^R/V_{cb}^L]\leq0.06$ is
present. 

\section{MSSM renormalization of the quark-quark-W vertex}

\begin{figure}
\includegraphics[width=0.45\textwidth]{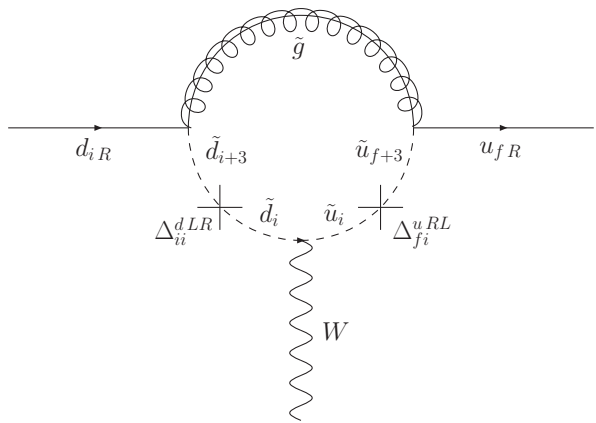}

\caption{ \label{RW}Feynman diagram which induces the effective right-handed W coupling of an down-type quark of flavour i to a up-type quark of flavor f. The crosses stand for the flavor and chirality changes needed to generate the coupling.}
\end{figure}

In Ref.~\cite{Crivellin:2008mq} the renormalization of the quark-quark-W
vertex by chirally enhanced supersymmetric self-energies has been
computed. The results imply new bounds on the off-diagonal elements of
the squark mass matrices if t'Hooft's naturalness criterion is invoked.
In this section we extend the analysis of
\cite{Crivellin:2008mq} and calculate the leading contributions to the 
quark-quark-W vertex which decouple for $M_{\rm SUSY}\to \infty$.   
Using the conventions of Ref.~\cite{Crivellin:2008mq} we expand to
first oder in the external momenta and decompose the self-energies as
\begin{eqnarray}
  \Sigma _{fi}^q  &=& \;\;\;\;\left( {\Sigma _{fi}^{q\;LR}  + \ps\Sigma
      _{fi}^{q\;RR} } \right)P_R \nonumber \\
&&\,+ \left( {\Sigma _{fi}^{q\;RL}  +
      \ps\Sigma _{fi}^{q\;LL} } \right)P_L . 
\end{eqnarray}
These self-energies lead to a flavor-valued wave-function
renormalization $\Delta U_{fi}^{q\;L,R}$ for all external left- and
right-handed fields. It is useful to decompose these factors further in
to an unphysical anti-Hermitian part $\Delta U_{fi}^{q\;L\;A}$, which can
be absorbed into the renormalization of the CKM matrix, and a Hermitian
part $\Delta U_{fi}^{q\;L\;H}$, which can constitute a physical effect 
appearing as a deviation from CKM unitarity: $\Delta U_{fi}^{q\;L,R\;H}  = \Sigma _{fi}^{q\;LL,RR}/2$. 
Neglecting external momenta, the genuine vertex-correction originating
from a squark-gluino loop is given by
\begin{widetext}
\begin{equation}
 - i\Lambda _{u_f d_i }^{W\;\tilde g}  \!=\! \frac{{g_2 }}{{\sqrt 2 }}\frac{{i\alpha _s }}{{3\pi }}\gamma ^\mu \!\! \sum\limits_{s,t = 1}^6 {\sum\limits_{j,k = 1}^3 \!{\left( {W_{fs}^{\tilde u} W_{ks}^{\tilde u*} V_{kj}^{L} W_{jt}^{\tilde d} W_{it}^{\tilde d *} P_L  + W_{f + 3,s}^{\tilde u} W_{ks}^{\tilde u *} V_{kj}^{L} W_{jt}^{\tilde d} W_{i + 3,t}^{\tilde d*} P_R } \right)} C_2 \left( {m_{\tilde u_s } ,m_{\tilde d_t } ,m_{\tilde g} } \right)} 
 \label{vertexkorrktur} .
\end{equation}
\end{widetext}
The matrices $W_{st}^{\tilde q}$ diagonalize the squark mass matrices
\cite{Crivellin:2008mq}. The part proportional to $P_L$ in \eq{vertexkorrktur} cancels
with the anti-Hermitian part of the wave-function renormalization
due to the SU(2) relation between the left-handed up and down
squarks for $M_{\rm SUSY}\to \infty$ according to the
decoupling theorem \cite{Appelquist:1974tg}. Since the loop functions
depend only weakly on $M_{\rm SUSY}$, the cancellation is very
efficient, even for light squarks around 300$\,$GeV. Therefore, the unitarity of the CKM matrix is conserved with very high accuracy.
A right-handed coupling of quarks to the W boson is induced by the diagram in Fig.~\ref{RW} if
left-right mixing of squarks is present. The effective coupling
corresponds to $Q_{RR}$ in \eq{Operator} and vanishes in the
decoupling limit. There is no wave-function renormalization of right-handed
quarks which can be applied to the W vertex, therefore no gauge cancellations occur.

\begin{widetext}

\begin{figure}
\includegraphics[width=0.4\textwidth]{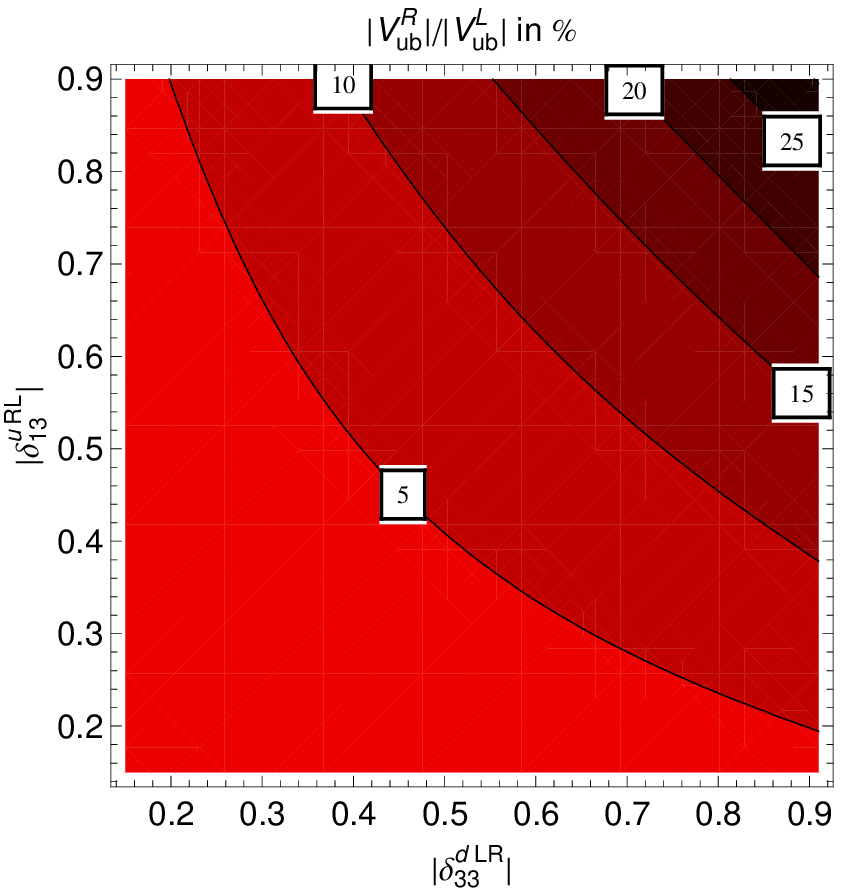}
\includegraphics[width=0.4\textwidth]{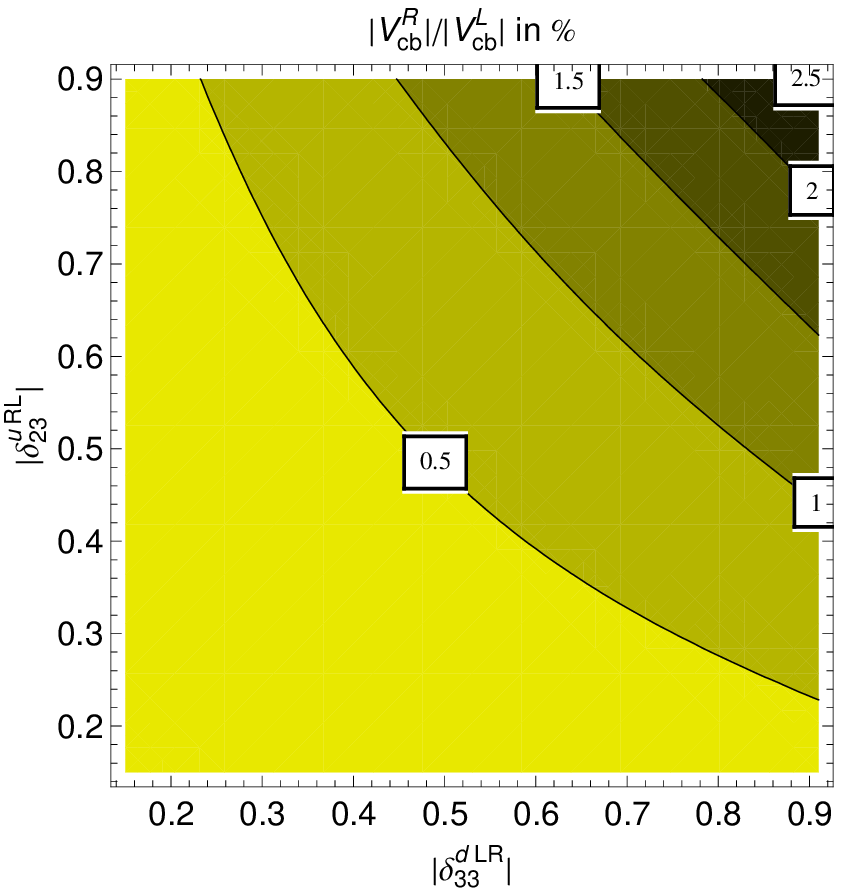}
\caption{\label{rvubcb} Left: Relative strength of the induced right-handed
  coupling $|V^R_{ub}|$ with respect to $|V^L_{ub}|$
  for $M_{\rm{SUSY}}=1\,\rm{TeV}$. $|V^L_{ub}|$ is determined from CKM unitarity. Right: Same as the left figure for $V_{cb}^R$.}
  \end{figure}
  
\end{widetext}

We show the relative size of the right-handed coupling involving u,c and b in Fig.~\ref{rvubcb}.
Note that the mass insertion $\delta^{u\;RL}_{13,23}$ are not affected by the fine-tuning
argument imposed in \cite{Crivellin:2008mq} nor severely restricted by
FCNC processes \cite{Dittmaier:2007uw}. Therefore the size of the induced couplings
$V^R_{ub}$ ($V^R_{cb}$) can be large enough to explain
(attenuate) the apparent discrepancies among the various determinations
of $|V_{ub}|$ ($|V_{cb}|$). Nevertheless, if $\delta^{u\;RL}_{13,23}$ is large single-top production is enhanced which can be observed at the LHC \cite{Plehn:2009it}. In principle also charged Higgs contributions to $B\to\tau\nu$ have to be considered in the MSSM. However, these contributions are only important in the special case in which both $\tan(\beta)$ is large and the charged Higgs is light. Furthermore, a charged Higgs always interferes destructively with the SM, making the discrepancy between the different determinations of $V_{ub}$ even bigger.
\section{Conclusions}
In this article we have first examined the effect of an
effective right-handed coupling of quarks to the W boson
on the determination of $|V_{ub}|$ and $|V_{cb}|$ from different decay
modes. In both cases a right-handed coupling can improve the agreement among these
determinations (Figs.~\ref{Vub} and \ref{Vcb}). 
In particular, one can
simultaneously remove the disturbing problem with $B\to\tau\nu_\tau$
\cite{Charles:2004jd} and improve the agreement among inclusive and
exclusive determinations of $|V_{ub}|$.
Second, we have shown that a
loop-induced right-handed coupling is generated within the MSSM if
left-right mixing of squarks is present.  This 
coupling has the right size needed to resolve the tensions
in $|V_{ub}|$. Such a scenario involves a large left-right mixing
between sbottoms (as present in e.g.\ the popular large-$\tan\beta$
scenarios) and a large $A^u_{31}$-term which enhances single-top production, making it observable at the LHC. If $\delta^{u\;RL}_{13}\approx0.6$ a 95\% CL signal can already be detected with 50 inverse femto-barn \cite{Plehn:2009it}. 
In $b\to c$ transitions the loop-induced supersymmetric right-handed coupling can alleviate, but
cannot fully remove, the discrepancies among the three methods to
determine $|V_{cb}|$. To probe $b\to u$ transitions we propose to look
for right-handed couplings in the differential decay distributions of
$B\to \rho \ell \nu_\ell$. The smaller right-handed component in 
$b\to c$ transitions can be probably better studied in $B\to X_c \ell\nu$
\cite{Dassinger:2008as} than in $B\to D^* \ell \nu$ decays, because 
a theoretical control of form factors to percent accuracy is
challenging.

{\it Acknowledgments.}---I like to thank Lars Hofer, Mikolaj Misiak, and Ulrich Nierste for
helpful discussions and proofreading the article. I also thank Frank Tackmann for bringing an error in the online version of "The Review of Particle Physics", which led to a mistake in first version of this article, to my attention.
This work is supported by BMBF grants 05HT6VKB and 
05H09VKF and by the EU Contract No.~MRTN-CT-2006-035482, 
\lq\lq FLAVIAnet''. I acknowledge the financial support of the State of
Baden-W\"urttemberg through \emph{Strukturiertes Promotionskolleg
Elementarteilchenphysik und Astroteilchenphysik}.

\bibliography{r-ckm}

\end{document}